\documentclass{appolb}
\usepackage{graphicx}
\begin{document}
\title{
Measurement of hadronic cross section \\at KLOE/KLOE-2
%
%\thanks{Presented at ...}%
}
\author{Veronica De Leo
\address{University of Messina, Italy}
\\
\address{on behalf of the KLOE/KLOE-2 Collaboration}
}
\maketitle
\begin{abstract}
%The measurement of the $\sigma(e^+e^-\rightarrow \pi^+\pi^-)$ cross section , obtained from the ratio 
%$\sigma(e^+e^-\rightarrow \pi^+\pi^-\gamma)/\sigma(e^+e^-\rightarrow\mu^+\mu^-\gamma$) has been performed with the KLOE detector at $DA\Phi NE$, the Frascati $\phi$-factory, using
%events with initial state radiation photons emitted at small angle and inclusive of the final state radiation. 
The measurement of the $\sigma(e^+e^-\rightarrow \pi^+\pi^-)$ cross section allows to determine the pion form factor $\vert F_\pi  \vert^2$ and the two pion contribution to the muon anomaly $a_\mu$. 
Such a measurement has been performed with the KLOE detector at $DA\Phi NE$, the Frascati $\phi$-factory.\\
%A significant reduction of the systematic error was reached respect to the previous KLOE measurements.   
%In 2008 and 2010 KLOE published two measurements of the $\pi^+\pi^−$ cross section;the first with the photon emitted at small angle, and the second with the
%photon emitted at large angle using data at the collision energy of 1 GeV.
%(i.e. 20 MeV below
%the $\phi$ -peak). While the measurements were normalized to the $DA\Phi NE$
%luminosity using large angle Bhabha scattering,The preliminary results on the combination of the last analysis (KLOE12) with the previous published (KLOE08,KLOE10) will be also presented.  
The preliminary results on the combination of the last analysis (KLOE12) with two previous published (KLOE08, KLOE10) will be presented in the following.  
%The result confirms the previous KLOE measurements and the current discrepancy ($\sim 3\sigma$) between the Standard Model (SM) calculation
%and the experimental value of the muon anomaly $a_\mu$.

\end{abstract}
\PACS{PACS numbers come here}
  
\section{Introduction}
The anomalous magnetic moment of the muon defined as $a_\mu \equiv \frac{g_\mu-2}{2}$,
can be accurately measured and, within the SM framework, precisely predicted\cite{Beringer}.
\,The experimental value of $a_\mu$ ($(11 659 208.9 \pm 6.3) \times 10^{-10}$)
measured at the Brookhaven Laboratory differs from the SM estimates by 3.2 - 3.6 $\sigma$ \cite{Bennett}. 
%This deviation in $a_\mu^{exp}$ from the SM expectation
%would signal effects of new physics.
If the deviation is confirmed with higher precision it would signal of new physics.% \cite{Czarnecki,Davier3}. 
%with current sensitivity
%reaching up to mass scales of \textit{O(TeV)} 
%reaching mass scales of \textit{O(TeV)} with current sensitivity\cite{Czarnecki,Davier3}. 
%The SM prediction for $a_\mu^{SM}$ can be distinguished into three
%parts: 
%\begin{equation}
%a_\mu^{SM}=a_\mu^{QED}+a_\mu^{EW}+a_\mu^{Had}
%\end{equation}

%\begin{figure}[h]
%\centering
%\includegraphics[width=7cm]{contrib_amu.jpeg}
%\caption{Representative diagrams contributing to $a_\mu^{SM}$. From left to right: first order QED
%(Schwinger term), lowest-order electro-weak, lowest order hadronic.}
%\label{contrib_amu.jpeg}       % Give a unique label
%\end{figure}

%the QED part that includes all photonic and leptonic (e, $\mu$, $\tau$) loops
%starting with the classic $\alpha/2\pi$ Schwinger contribution; 
%the loop contributions involving heavy $W\pm$, $Z$ or Higgs particles
%that are collectively labeled as $a_\mu^{EW}$ and finally the  
%hadronic (quark and gluon) loop contributions.  
%The hadronic contribution, which is the responsible of the main theoretical uncertainties, is given 
%itself by different contributions:

%\begin{equation}
%a_\mu^{Had}=a_\mu^{Had,LO}+a_\mu^{Had,HO}+a_\mu^{Had,LBL}
%\end{equation} 

%where $a_\mu^{Had,LO}$ is the lowest-order contribution from hadronic
%vacuum polarization, $a_\mu^{Had,HO}$ is the corresponding higher-order
%part and the last term $a_\mu^{Had,LBL}$ is the light-by-light
%(LBL) scattering part \cite{Davier3}.
%The hadronic contribution to $a_\mu$ is the responsible of the main theoretical uncertainties.
The main theoretical uncertainty for  $a_\mu$ comes from hadronic contributions.
The leading order hadronic term can be derived from 
a combination of experimental cross section data, 
%involving
%$e^+e^-$ annihilation to hadrons, and perturbative QCD. These
%are used to evaluate an energy-squared dispersion integral:
related to $e^+e^-$ annihilation to hadrons.
 %and perturbative QCD. %These terms are used to evaluate an energy-squared dispersion integral:
%\begin{equation}
%a_\mu^{Had}[LO]= \frac{1}{3} \left(\frac{a}{\pi}\right)^2 \int_{{m_\pi^2}}^\infty ds \frac{K(s)}{s}R(s)
%\end{equation}

%where $K(s)$ is a QED kernel function \cite{De Rafael}, and where $R(s)$
%denotes the ratio of the cross section for $e^+e^-$ annihilation
%into hadrons to the pointlike muon-pair cross section at center of
%mass energy $\sqrt{s}$. \\
%The integration
%kernels occurring in the dispersion relations emphasise low
%photon virtualities, 
%due to the 1/s slope of the cross section.\\
At $DA\Phi NE$
the differential cross section (as a function of $m_{\pi\pi}$) for the $e^+e^-\rightarrow \pi^+\pi^-\gamma$ 
%as a function of the $\pi^+\pi^-$ invariant mass, $M_{\pi\pi}$ , 
 initial state radiation (ISR) 
process is measured. Then, the dipion cross section $\sigma_{\pi\pi} \equiv \sigma (e^+e^- \rightarrow \pi^+\pi^-)$ has been obtained from:
\begin{equation}
s \frac{d\sigma(\pi^+\pi^-\gamma)}{ds_\pi} \vert_{ISR} = \sigma_{\pi\pi}(s_\pi) H(s_\pi ,s),
\label{dipion_cross_section} 
\end{equation}

where the radiator function \textit{H} is computed from QED with complete
NLO corrections %\cite{Rodrigo,Czyz,Czyz1,Czyz2,Actis} 
and depends on the initial $e^+e^-$ center of mass energy squared $s$. The dipion cross section 
$\sigma_{\pi\pi}$ obtained from Eq. \ref{dipion_cross_section} requires the correction
for final state radiation (FSR).
Eq. \ref{dipion_cross_section} is also valid for the $e^+e^- \rightarrow \mu^+\mu^-\gamma$ and $e^+e^- \rightarrow \mu^+\mu^-$ processes
with the same radiator function \textit{H}. Thus, we can determine  
$\sigma_{\pi\pi}$ from the ratio of the $\pi^+\pi^-\gamma$ and $\mu^+\mu^-\gamma$ differential cross
sections for the same value of the dipion and dimuon invariant
mass. \\%(see also Refs. \cite{Aubert,Lees}).\\
The pion form factor can then be determined using the following equation:
\begin{equation}
{\vert F_\pi(s') \vert}^2 = \frac{3}{\pi} \frac{s'}{\alpha^2 \beta_\pi^3} \sigma^0_{\pi\pi(\gamma)}(s')
(1+\delta_{VP}) (1-\eta_\pi(s'))
\label{F_pi}
\end{equation}

where $\delta_{VP}$ is the Vacuum Polarization (VP) correction %\cite{Fred1}
, $\eta_\pi$ accounts for the FSR radiation
assuming point-like pions. 
%\cite{Schwinger}
$\sigma^0_{\pi\pi}$ is a bare cross section, i.e. corrected for the
running of $\alpha_{em}$ and inclusive of FSR, defined as \cite{Babusci}

$\sigma^0(\pi^+\pi^-,s')=$
\begin{equation}
\frac{d\sigma(\pi^+\pi^-\gamma,ISR)/ds'}{d\sigma(\mu^+\mu^-\gamma,ISR)/ds'}
\times \sigma^0(e^+e^-\rightarrow\mu^+\mu^-\gamma,s')
\end{equation}
where $s'=s_\pi=s_\mu$. \\
%Many radiative corrections drop out using the ratio method as the radiator function (so, the measurement of $\sigma_{\pi\pi}$ is not affected by the related systematic uncertainty of $0.5\%$), 
%the integrated luminosity (since the data sample for the $\pi^+\pi^-\gamma$ and $\mu^+\mu^-\gamma$ events are the same) and the vacuum polarization.
Many radiative corrections drop out for this ratio method: contributions due to the radiator function (this allows to suppress the related systematic
uncertainty of 0.5\% for the direct $\sigma_{\pi\pi}$ measurement), to the integrated luminosity
(since the data for the $\pi^+\pi^-\gamma$ and $\mu^+\mu^-\gamma$ processes are collected simultaneously) and
finally to the vacuum polarization.
%\section{KLOE Detector}
%The KLOE detector operates at $DA\Phi NE$, the Frascati $\phi$-factory,
%an $e^+e^-$ collider running at fixed energy, $W =\sqrt{s} \sim 1020 MeV$,
%the $\phi$ meson mass.\\
%\begin{figure}[h]
%\centering
%\includegraphics[width=4.5cm]{Kloe_detector1.jpeg}
%\caption{Schematic view of the KLOE detector. In the picture the superconducting coil (yellow), 
%the electromagnetic calorimeter (red) and the cylindrical drift chamber (blue) are visible.}
%\label{detector}      
%\end{figure}
\section{Measurement of the $e^+e^- \rightarrow \pi^+\pi^-$ cross section at KLOE}
\begin{figure}[h]
\centering
\includegraphics[width=4.3cm]{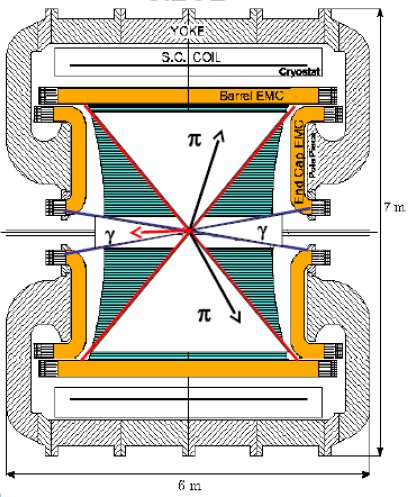}
\caption{Schematic view of the KLOE detector with selection regions.}
\label{detector08}      
\end{figure}

In the 2008 and 2010 two  analyses of the $\sigma(e^+e^- \rightarrow \pi^+\pi^-\gamma)$ have been 
performed at $DA\Phi NE$ with the KLOE detector. \\
%It consists of a cylindrical drift chamber (DC)
%\cite{Adinolfi} and an electromagnetic calorimeter (EMC) \cite{Adinolfi1}. The DC has a
%momentum resolution of $\sigma_{p_\perp}/p_\perp \sim 0.4\%$ for tracks with polar angle
%$\theta > 45^\circ$. Track points are measured in the DC with a resolution
%in $r-\phi$ of $\sim$ 0.15 mm and $\sim$ 2 mm in z. The EMC has an energy
%resolution of $\sigma_E/E \sim 5.7\%/$ 
%$\sqrt E(GeV)$ and an excellent time resolution
%of $\sigma_t \sim 54 ps/\sqrt E (GeV) \oplus 100 ps$. A superconducting coil provides an axial magnetic field of
%0.52 T along the bisector of the colliding beam directions. 
A cross section of the detector in
the y, z plane is shown in Fig.\ref{detector08}.\\
The KLOE08 analysis \cite{Ambrosino} used a data sample corresponding to an integrated luminosity of
$240$ pb$^{-1}$ collected at $\sqrt{s}=m_\phi$ in 2002 and 
selection cuts in which 
the photon is emitted within a cone of $\theta_\gamma < 15^\circ$ around the beamline (narrow cones in Fig.\ref{detector08}) and the two charged pion tracks have $50^\circ < \theta_\pi < 130^\circ$ (wide cones in Fig. \ref{detector08}). In this
configuration, the photon is not detected and the photon momentum is reconstructed from missing momentum: $ \vec{p_\gamma}\simeq \vec{p_{miss}} = -(\vec{p_+}+\vec{p_-})$. These selection cuts provide high statistics data sample for the ISR signal events, and significantly reduce contamination from the resonant process $e^+e^- \rightarrow \phi \rightarrow \pi^+\pi^-\pi^0$.

%Using Eq. \ref{dipion_cross_section}
%the pion form factor $\vert F_\pi \vert^2$ is extracted.
From the bare cross section, %i.e. corrected for the
%running of $\alpha_{em}$ and inclusive of FSR, 
the dipion contribution to the muon anomaly $\Delta^{\pi\pi}a_\mu$ is measured:

$\Delta^{\pi\pi}a_\mu (0.592<M_{\pi\pi}<0.975GeV)=(387.2 \pm 3.3)\times 10^{-10}$.

The KLOE10 analysis \cite{Ambrosino10} was performed requiring events that are selected to have a photon at large polar
angles between $50^\circ < \theta_\gamma < 130^\circ$ (wide cones in Fig.\ref{detector08}), in the same angular region as the pions.\\
This selection allow to access the two pion threshold. 
However, compared to the measurement with photons at small angles, this condition reduces statistics 
%of about a factor 5
and increases the background from the process 
$\phi \rightarrow \pi^+\pi^-\pi^0$. 
The dispersion integral for $\Delta^{\pi\pi}a_\mu$ is computed as the sum of
the values for $\sigma^0_{\pi\pi(\gamma)}$ times the kernel K(s), times
$\Delta s = 0.01 GeV^2$ %\cite{Babusci}

\begin{equation}
\Delta^{\pi\pi}a_\mu = \frac{1}{4\pi^3}\int^{s_{max}}_{s_{min}} ds \, \sigma^0_{\pi\pi(\gamma)}(s)K(s).
\label{delta_amu}
\end{equation}

The following value for the dipion contribution to the muon anomaly $\Delta^{\pi\pi}a_\mu$ was found:

$\Delta^{\pi\pi}a_\mu (0.1-0.85) GeV^2 = (478.5 \pm 2.0_{stat} \pm 5.0_{exp} \pm 4.5_{th})\times 10^{-10}$.

%In the fig. \ref{08_10} the comparison between the measurements of the  pion form factor of the KLOE08 and KLOE10 analysis is reported. \\ 
%As one can see, an excellent agreement is found for $(M^0_{\pi\pi})^2 > 0.5 GeV^2$ , while below the KLOE10 result is lower by few percent.
%We stress that the data sets have been obtained at different operating conditions of the $DA\Phi NE$ collider, and different selection cuts in acceptance were used. Also the
%analysis procedures were different since in the KLOE08 analysis
%the radiated photon was not detected.
%uncomment the following lines to place a figure
%\begin{figure}[htb]
%\centerline{%
%\includegraphics[width=12.5cm]{Fig1}}
%\caption{Plot of ...}
%\label{Fig:F2H}
%\end{figure}
%\section{Measurement of the pion form factor from the $\pi\pi\gamma /\mu\mu\gamma$ ratio}
The last KLOE measurement of the $e^+e^- \rightarrow \pi^+\pi^-$ cross section (KLOE12) has been 
obtained from the ratio between the pion and muon ISR differential cross section.   
The data sample is the same as for the KLOE08 analysis. \\%and corresponds to an integrated luminosity of
%$240 pb^{-1}$ collected in 2002.\\
The separation between the $\pi\pi\gamma$ and $\mu\mu\gamma$ events is obtained 
assuming the final state with two charged particles
with equal mass $M_{TRK}$ and one photon. The $M_{TRK} < 115 MeV$ identifies the muons and $M_{TRK} > 130 MeV$ the pions. 
%It is computed from the energy and momentum conservation laws:\\
%From the energy and momentum conservation laws we obtain:
%\begin{equation}
%(\sqrt{s}-\sqrt{\vert \vec{p_+}\vert^2 +M_{TRK}^2}-\sqrt{\vert \vec{p_-}\vert^2 +M_{TRK}^2})^2-(\vec{p_+}+\vec{p_-})^2=0
%\end{equation}

%where $\vec{p_\pm}$ is the measured momentum of the positive (negative) particle.

%\begin{figure}[h]
%\centering
%\includegraphics[width=3.8cm]{m_trk.jpeg}
%\caption{Separation of pion and muon events by cuts in $M_{TRK}$.}
%\label{m_trk}      
%\end{figure}  

%This selection leads to $8.9 \times 10^5 \mu\mu\gamma$ events and about $34.9 \times
%10^5$ for $\pi\pi\gamma$ events.
%It is checked against other techniques, 
The selection procedure has
been compared to other techniques,
such as a kinematic fit or applying a quality cut on the helix fit for both tracks,
%tighter cuts on the quality of the charged tracks,
 all leading
to consistent results.\\
Trigger, particle identification and tracking efficiencies have been checked using
control data samples.\\
The differential $\mu\mu\gamma$ cross section is obtained from the observed
event count $N_{obs}$ and background estimate $N_{bkg}$, as: 

\begin{equation}
\frac{d\sigma_{\mu\mu\gamma}}{ds_\mu}= \frac{N_{obs}-N_{bkg}}{\Delta{s_\mu}}\frac{1}{\epsilon(s_\mu)\textit{L}}
\end{equation}

where \textit{L} is the integrated luminosity from Ref. \cite{Ambrosino3} and $\epsilon(s_\mu)$ the
selection efficiency. 
The bare cross section $\sigma^0_{\pi\pi(\gamma)}$
(inclusive of FSR, with VP effects removed) is obtained from the bin-by-bin
ratio of the KLOE08 $\pi\pi\gamma$  and the described above $\mu\mu\gamma$ differential cross sections. The bare cross section is used in the dispersion integral to
compute $\Delta^{\pi\pi}a_\mu$.
The pion form factor $\vert F_\pi \vert^2$ is extracted using Eq. (\ref{F_pi}).\\ 
Eq. \ref{delta_amu} gives $\Delta^{\pi\pi}a_\mu  =
(385.1 \pm 1.1_{stat} \pm 2.6_{exp} \pm 0.8_{th}) \times 10^{-10}$ in the interval 
$0.35 <M_{\pi\pi}^2 < 0.95 GeV^2$. For each bin contributing to the integral, statistical
errors are combined in quadrature and systematic errors
are added linearly.\\
The last three KLOE estimations on the $\Delta^{\pi\pi}a_\mu$ (KLOE08, KLOE10, KLOE12) have been compared and are consistent (as you can see in the table \ref{tab1}). 
\begin{table}[h]
\centering
\caption{Comparison of $\Delta^{\pi\pi}a_\mu$  between the KLOE12 and the previous
KLOE measurements (KLOE08, KLOE10).}
\label{tab1}       % Give a unique label
% For LaTeX tables you can use
\begin{tabular}{lll}
\hline
Measurement  & $\Delta a_\mu^{\pi\pi}(0.35-0.95 GeV^2)\times 10^{10}$ &   \\\hline
KLOE12 & $385.1 \pm 1.1_{stat} \pm 2.7_{sys+theo}$ &  \\
KLOE08 & $387.2 \pm 0.5_{stat} \pm 3.3_{sys+theo}$ &  \\\hline
 & $\Delta a_\mu^{\pi\pi}(0.35-0.85 GeV^2)\times 10^{10}$ &   \\\hline
KLOE12 & $377.4 \pm 1.1_{stat} \pm 2.7_{sys+theo}$ &  \\
KLOE10 & $376.6 \pm 0.9_{stat} \pm 3.3_{sys+theo}$ &  \\
\end{tabular}
\end{table}
\begin{figure}[h!]
%\centering
%\includegraphics[width=9.3cm]{combination1.jpeg}
%\includegraphics[width=5.6cm]{comb.jpeg}
\includegraphics[width=6cm]{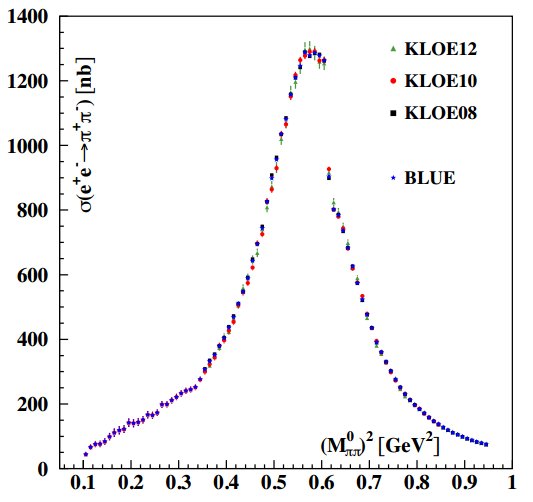}
\includegraphics[width=6cm]{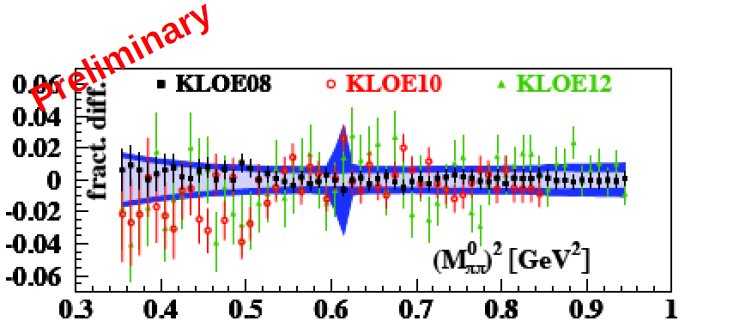}
%\hspace{5cm}
\begin{center}
\textbf{$\frac{\vert F_{KLOEXX}\vert^2-\vert F_{BLUE}\vert^2}{\vert F_{BLUE}\vert^2}$}\\
%\vspace{0.2cm}
%\includegraphics[width=7.5cm]{tab_KLOE.png}
\end{center}
\caption{Preliminary combination of the last three KLOE results (KLOE08, KLOE10, KLOE12) on the pion form factor measurements (left) and the fractional difference (right) using the Best Linear Unbiased Estimate (\textit{BLUE}) method \cite{Valassi,D'Agostini}. The light blue band in the fractional difference is the statistical error and the dark blue band is the combined statistical and systematic uncertainty\cite{Mueller}.}
\label{combination}
\end{figure}
 The preliminary combination of these KLOE results is reported in the figure \ref{combination}\cite{Mueller}.
It is obtained using the Best Linear Unbiased Estimate (\textit{BLUE}) method \cite{Valassi,D'Agostini}.
In the Fig. \ref{combination} (left) the pion form factor measurements for the three KLOE analysis and the fractional difference (right) are shown\cite{Mueller}. 
%The comparison between the last $\Delta^{\pi\pi}a_\mu$ KLOE measurement 
%and the previous KLOE08 and KLOE10 measurements is visible in the right part of the same figure. 
%region is found (as it's possible to see from the Fig.\ref{fpi_12_8.jpeg}).
%The Fig.8
%shows the good agreement between the KLOE12 and KLOE08 measurements, in
%particular in the $\rho$ region.
The cross section ratio method used in the KLOE12 measurement reduces significantly the theoretical  and the systematic error.\\
The following $a^\mu_{\pi\pi}$ values are found:\\

$a^\mu_{\pi\pi}(0.1-0.95GeV^2)= (487.8 \pm 5.7)\cdot 10^{-10}$

$a^\mu_{\pi\pi}(0.1-0.85GeV^2)= (378.1 \pm 2.8)\cdot 10^{-10}$.
%\vspace{0.4cm}
\section{Conclusion}
Precision
measurements of the pion vector form factor using the Initial State Radiation (ISR) have been performed by the KLOE/KLOE-2 Collaboration during the last 10 years. % which confirmed a $3\sigma$ discrepancy between $a_\mu$ SM and
%the value measured at BNL.
 The preliminary consolidation of the last analysis (KLOE12) with two previously published (KLOE08, KLOE10) ones has been presented.  
%The preliminary results on the combination of the last analysis (KLOE12) with two previous published (KLOE08, KLOE10) has been presented.  
The result confirms the current discrepancy ($\sim 3\sigma$) between the Standard Model (SM) calculation
and the experimental value of the muon anomaly $a_\mu$ measured at BNL.\\
In the near future the $\gamma \gamma$ Physics program of the KLOE-2 
experiment\cite{kloe2physics}
will further shed light in this field, with e.g. the study of the 
radiative width of pseudoscalar mesons
and of the transition form factors \cite{pi0gg},
thanks to the luminosity upgrade of DA$\phi$NE
and the KLOE upgrade with the addition of new detectors:
low energy \cite{let} and high energy \cite{het} $e^+e^-$ taggers, an
inner tracker \cite{kloeIT}, crystal calorimeters (CCALT) \cite{ref1}, 
and tile calorimeters (QCALT) \cite{ref2}.

%The published measurements (KLOE08, KLOE10), normalized to Bhabha
%events, have allowed the measurement of $a_\mu^{\pi\pi}$
%in the region below 1 GeV with $\sim1\%$ total
%error.
%A new measurement (KLOE12) of  $\vert F_\pi \vert^2$from the $\pi\pi\gamma/\mu\mu\gamma$ ratio (based on 240 $pb^{-1}$)
%with $0.7\%$ systematic error has been published \cite{Babusci}.
%This pion form factor determination is in very good agreement
%with previous KLOE results.

% end of file template.tex

\end{document}